\documentclass[aps,superscriptaddress,pra,twocolumn,footinbib]{revtex4-1}
\usepackage{amsmath,amssymb}
\usepackage{bm}
\usepackage{epsfig}
\usepackage{natbib}
\usepackage{hyperref}
\usepackage{graphicx}
\usepackage{epstopdf}

 \hypersetup{pdfstartview={FitH},pdfpagemode={UseNone},
            colorlinks,linkcolor=red, citecolor=blue, urlcolor=blue,
            bookmarks=true, bookmarksopen=true, pdfnewwindow=true}
             
\usepackage{verbatim} 
\usepackage{morefloats}
\usepackage{bibentry}
\usepackage[mathscr]{euscript}

\newcommand{\eps}{\varepsilon}
\newcommand{\rmi}{{\rm i}}
\newcommand{\rmd}{{\rm d}}
\newcommand{\rot}{\mathop{\mathrm{rot}}\nolimits}

\renewcommand{\Im}{\mathop{\mathrm{Im}}\nolimits}
\renewcommand{\Re}{\mathop{\mathrm{Re}}\nolimits}

\renewcommand{\phi}{\varphi}

\begin{document}

\title{
Nonlinear generation of quantum-entangled photons from \\
high-$Q$ states in dielectric nanoparticles
}

\author{\firstname{Alexander N.} \surname{Poddubny}}
\email{poddubny@coherent.ioffe.ru}
\affiliation{Ioffe  Institute, St.~Petersburg 194021, Russia}
\affiliation{Nonlinear Physics Centre, Australian National University, Canberra ACT 2601, Australia}
\author{\firstname{Daria A.} \surname{Smirnova}}
\affiliation{Nonlinear Physics Centre, Australian National University, Canberra ACT 2601, Australia}
\affiliation{Institute of Applied Physics, Russian Academy of Science, Nizhny Novgorod 603950, Russia}

\begin{abstract}
We develop multipolar theory of nonlinear generation of entangled photons from subwavelength dielectric particles due to the spontaneous parametric downconversion. 
We demonstrate that optical excitation in resonance with the high-quality supercavity mode of the aluminum gallium arsenide (AlGaAs) nanodisk leads to a strong enhancement of generation of entangled photon pairs associated with electric and magnetic dipole modes. Our rigorous numerical results are corroborated by an analytical model, universally describing formation of high-$Q$ resonant states and dark states due to the interference and interplay of the parent multipoles, namely, magnetic dipoles and magnetic octupole. Our findings and description can be instructive for quantum and nonlinear nanophotonics applications.
\end{abstract}

\maketitle

\section{Introduction}\label{sec:intro}

Quantum communications and information processing technologies demand for  compact emitters of entangled photons~\cite{Silverstone2013,wabnitz2015all}. Despite the recent advances with the quantum dot setups~\cite{Somaschi2016,Reithmaier2015}, so far the spontaneous wave mixing in nonlinear crystals 
remains the most efficient approach for room-temperature  two-photon generation~\cite{Guo2016,Wang:14}. However, bringing efficient and integrable quantum nonlinear emitters to the nanoscale is quite challenging. A promising recently emerged versatile platform to enhance nonlinear processes
is based on semiconductor nanoparticles, such as Si and AlGaAs, supporting pronounced Mie-type resonances and possessing strong optical nonlinearities~\cite{Smirnova:16}. So far most of the studies have been focused on enhancement of classical nonlinear processes, including second-harmonic~\cite{Makarov2017,Smirnova2018}, sum-frequency and third harmonic~\cite{Shcherbakov2014,Morales2016,Smirnova2016} generation. While initial experiments on the  nonlinear entangled photon generation from  AlGaAs nanoantennas are already available~\cite{Marino,Marino2018b}, fundamental and complementary design approaches are still required to improve the modes quality and the resulting generation efficiency.

Recently it has been suggested to utilize so-called high-quality (high-$Q$) resonant states to enhance the simplest classical nonlinear process of second-harmonic generation in dielectric nanodisks~\cite{Koshelev2018}. These states are akin to the bound states of continuum  (BICs) in infinite periodic dielectric structures~\cite{Hsu2016}: their high finesse is due to the destructive interference of several far-field radiation channels~\cite{Rybin2017,Bogdanov2018}. However, the multipolar nature of such high-$Q$ resonant states and their potential for applications in quantum nanophotonics have not yet been studied.

  \begin{figure}  [b!]
\centering\includegraphics[width=0.45\textwidth]{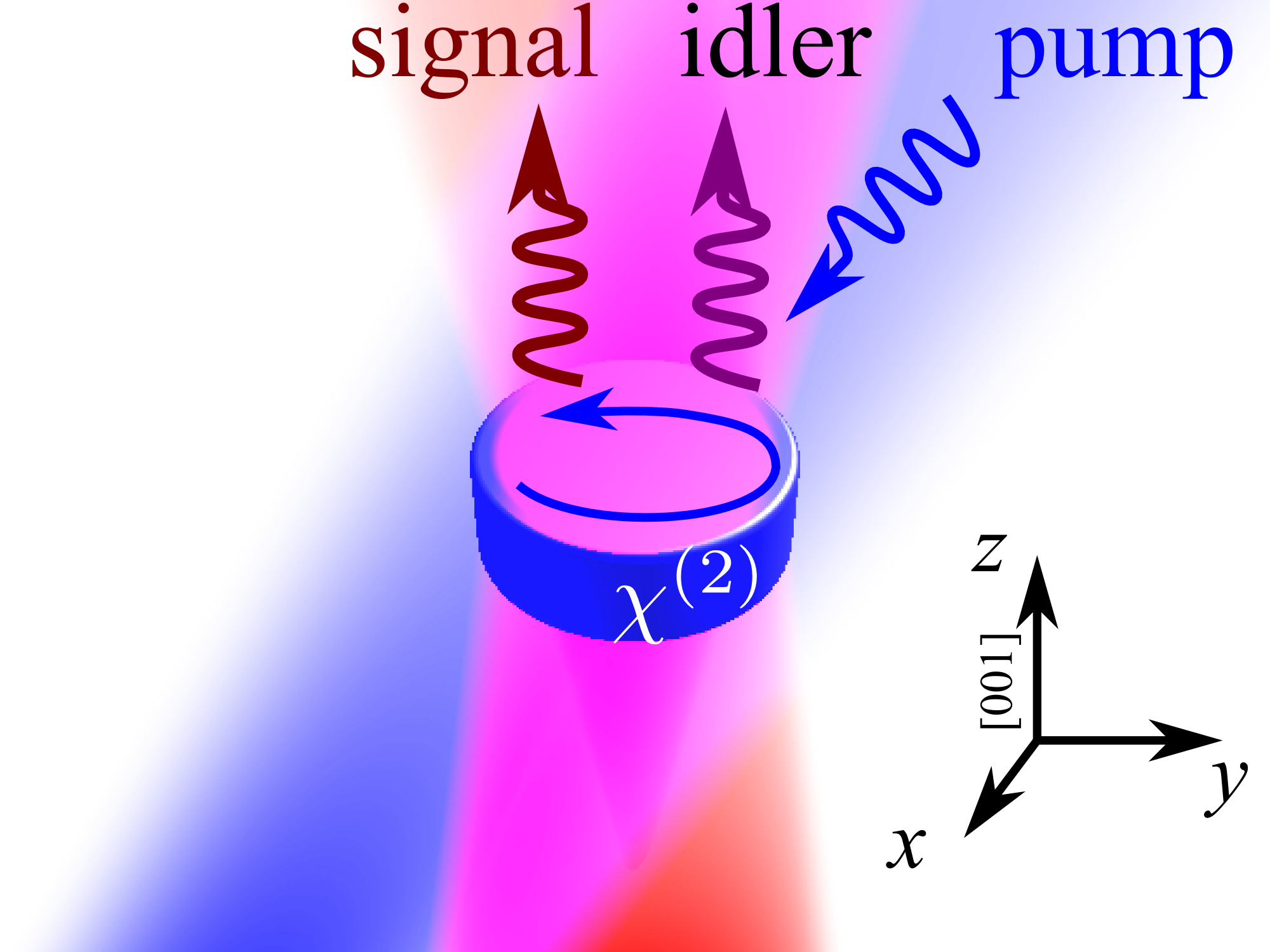}
 \caption{Schematic of the entangled photons generation from a nonlinear nanodisk.
TE-polarized plane wave pump excites a hybridized quasi-magnetic dipole-octupole high-Q state with angular momentum projection $M=0$. The signal and idler photons are in in-plane electric and magnetic quasi-dipole modes.}\label{fig:1}
 \end{figure}

Here we put forward the high-$Q$ resonant states for the enhanced generation of entangled photons by spontaneous parametric downconversion (SPDC) in the nanodisk with the bulk quadratic nonlinearity $\chi^{(2)}$. Our calculation reveals an enhancement of generation of entangled photons in electric and magnetic dipole modes when a sum of their frequencies is equal to that of the high-$Q$ state. Our model is based on the general rigorous Green-function-based quantum-optical theory for two-photon generation in arbitrary nonlinear nanostructures we recently developed~\cite{Poddubny2016SPDC,Lenzini2018}. We also propose a fully analytical 3-level model
 that  provides useful insights in the formation of resonant states. Explicit account for the two magnetic quasi-dipole modes and one magnetic quasi-octupole mode generalizes the previous 2-state approaches~\cite{Rybin2017,Bogdanov2018} and allows us a universal simultaneous description of high-$Q$ states and dark states in the multipolar scattering of dielectric nanodisks. It might also be useful to describe the characteristic Fano features in the optical scattering spectra~\cite{Bogdanov2018}.

\section{Linear multipolar scattering}\label{sec:model}
The scheme of the structure under consideration is shown in Fig.~\ref{fig:1}. We consider  a semiconductor nanodisk in air with the permittivity $\eps$ and nonlinear susceptibility $\chi^{(2)}$, oriented along the $z$ axis. Our numerical approach is based on the $t$-matrix or extended boundary condition method~\cite{Waterman1971,tsang2000scattering,mishchenko2002scattering}. Namely, the electric field is expanded over the vector spherical harmonics  in the following way:
\begin{equation}
\bm E(\bm r)=
\begin{cases}
\sum\limits_{\nu}[a_{0,\nu} \bm J_{\nu}(k_{\rm out}\bm r)+a_{{\rm scat},\nu} \bm H_{\nu}(k_{\rm out}\bm r)], & \text{(air)}\\
\sum\limits_{\nu}a_{{\rm in},\nu} \bm J_{\nu}(k_{\rm in}\bm r), & \text{(disk)}\:
\end{cases}
\end{equation}
outside and inside the disk, respectively. Here the index $\nu$ incorporates the  total angular momentum $J$, its projection to $z$ axis $M$ and the polarization $\sigma=$TE,TM; $k_{\rm out}=\omega/c$ and $k_{\rm in}=\omega\sqrt{\eps}/c$. The functions $\bm J$ and $\bm H$ are regular and outgoing vector spherical  harmonics~\cite{tsang2000scattering}. The scattering is described by the $Q$ and $t$ matrices, linking the incident field with the field inside the disk and the scattered field,
\begin{equation}
a_{\rm in}=Q^{-1}a_{0},\quad a_{\rm scat}=ta_{0}
\end{equation}
and related by
\begin{equation}
t=-\rmi \Im Q\cdot Q^{-1}\:.\label{eq:tq}
\end{equation}
While the numerical calculation of the $t$ matrix is straightforward, it is instructive to study analytically how the interference between different multipolar modes manifests in the scattering.
  \begin{figure}
\centering\includegraphics[width=0.4\textwidth]{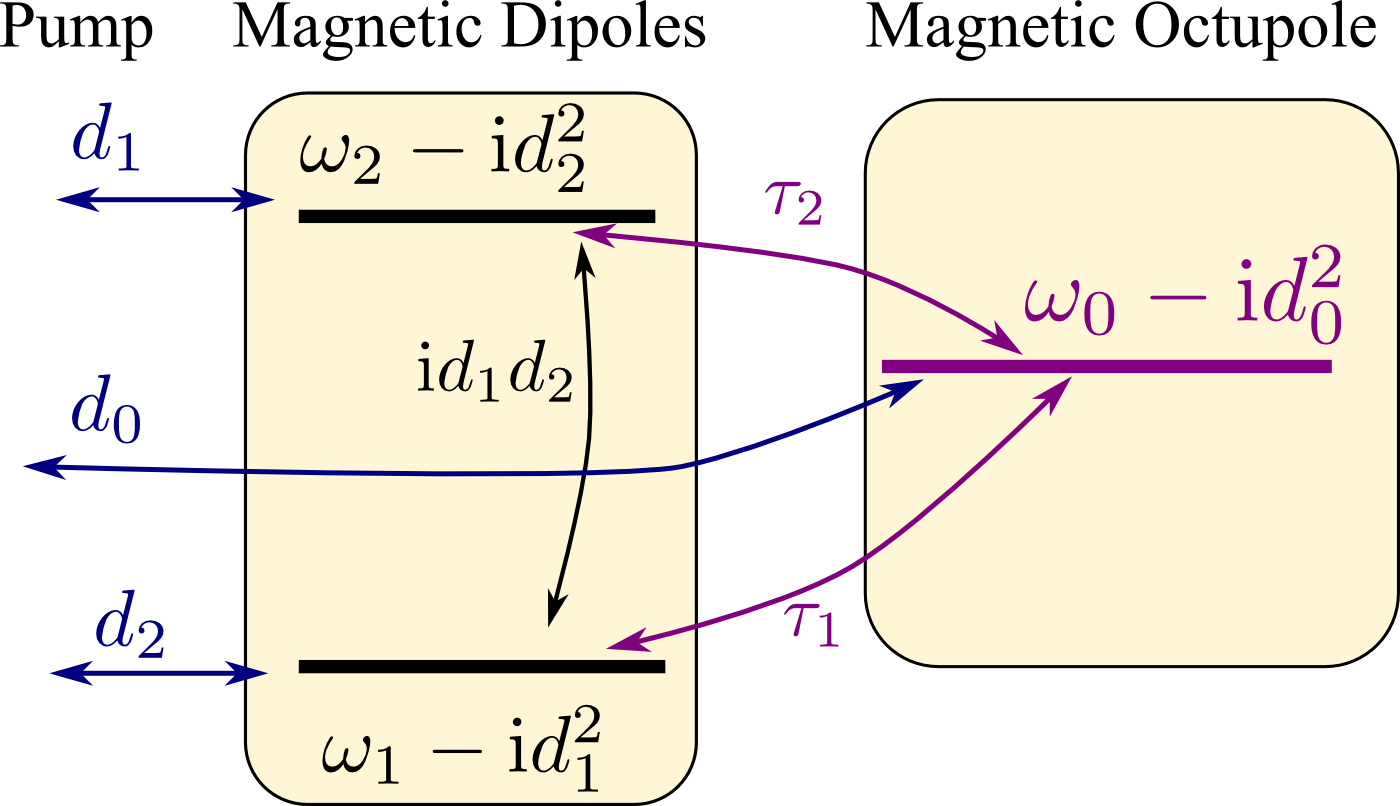}
 \caption{Schematic of the three-level model for the formation of dark and high-Q state, captured by the Hamiltonian Eq.~\eqref{eq:H}.}\label{fig:2}
 \end{figure}
\subsection*{Analytical 3-state model}\label{sec:an}
In the case of spherical symmetry the $t$-matrix is diagonal. For cylindrical symmetry it remains diagonal in angular momentum projection $M$ only, while the modes with different total momentum or polarization  can hybridize. Specifically, the hybridization takes place between the modes with the same polarization and even $J-J'$ as well as different polarization and odd $J-J'$.  The   high-$Q$ resonant state, proposed in \cite{Rybin2017}, corresponds to $M=0$, when the TE and TM polarizations are decoupled. In order to elucidate its nature we put forward an analytical 3-level model,
schematically  illustrated in Fig.~\ref{fig:2}. Our consideration generalizes the approach developed in Ref.~\cite{Hsu2014}, describing the dark state in scattering by a mutual suppression of two scattering channels. Here we also include a third state and the resulting basis includes two  quasi-magnetic-dipole states $|1\rangle,|2\rangle $ with $m=0$, and first and second radial quantum numbers and the lowest quasi-magnetic-octupole state $|0\rangle$.

The non-Hermitian Hamiltonian of the structure assumes the form
\begin{align}
 H=\begin{pmatrix}
\omega_{1}-\rmi d_{1}^{2}&-\rmi d_{1}d_{2}&\tau_{1}\\
-\rmi d_{1}d_{2}&\omega_{2}-\rmi d_{2}^{2}&\tau_{2}\\
\tau_{1}&\tau_{2}&\omega_{0}-\rmi d_{0}^{2}&\\
\end{pmatrix}\label{eq:H}\:.
\end{align}
The $d$ coefficients are the effective dipole moments, governing both the radiative decay of the states and their coupling to the external field. The $\tau$ coefficients describe the hybridization of the dipole and octupole modes.
Ohmic losses are neglected.
The $t$-matrix in the basis of dipole and octupole channels, $D\equiv J=1,M=0,TE$ and $O\equiv J=3,M=0,TE$, is given by
\begin{equation}
t=\rmi u(H-\omega)^{-1}u^{T},\quad u=\begin{pmatrix}
d_{1}&d_{2}&0\\
0&0&d_{0}
\end{pmatrix}\:,
\end{equation}
where the $u$ matrix describes the coupling of  the incident field to the states $0,1,2$. It is not quadratic since there exist two  dipole modes and only one octupole mode.
It is convenient to present the $t$-matrix in our model in the form Eq.~\eqref{eq:tq}, where the $Q$ matrix reads
\begin{align}\label{eq:q1}
\Re Q&=\begin{pmatrix}
-\dfrac{(\omega_{1}-\omega)(\omega_{2}-\omega)}{v_{DD}}&-\dfrac{v_{DD}}{d_{0}v_{DO}}
\\
-\dfrac{v_{DO}}
{d_{O}v_{DD}}
&
\dfrac{\omega_{0}-\omega}{d_{0}^{2}}
\end{pmatrix}\:,\\
\Im Q&=\begin{pmatrix}
1&\dfrac{(d_{1}t_{2}-d_{2}t_{1})^{2}}{d_{0}v_{DO}}\\0&1
\end{pmatrix}\:,\\
v_{DD}&=(\omega_{1}-\omega)d_{2}^{2}+
(\omega_{2}-\omega)d_{1}^{2},\label{eq:q3}\\
v_{DO}&=d_{1}t_{1}(\omega_{2}-\omega)+d_{2}t_{2}(\omega_{1}-\omega)\:.\label{eq:q4}
\end{align}
An explicit form of the $t$-matrix can be readily calculated from Eqs.~\eqref{eq:q1}--\eqref{eq:q4}. It  fully satisfies the reciprocity and unitarity conditions $t^{T}=t, S^{\dag}S=1$ where $S=1+2t$ is the scattering matrix. The advantage of the form 
 Eqs.~\eqref{eq:q1}--\eqref{eq:q4} is that it facilitates the extraction of the model parameters from the numerically computed $Q$-matrix. In that follows, we are most interested in the imaginary part of the eigenfrequency of the high-$Q$-state, that can be found from the position of the resonance of  $t_{OO}$ and reads
\begin{equation}
-\Im \omega_{O}=d_{0}^{2}+\frac{v_{DO}^{2}}{v_{DD}^{2}+(\omega_{1}-\omega)^{2}(\omega_{2}-\omega)^{2}}\:.\label{eq:wO}
\end{equation}
Two essential results of the model Eq.~\eqref{eq:q1}--\eqref{eq:wO} are as follows.
First, the  dark state is formed in the dipole channel due to the destructive interference of first and second dipole modes in the far field, i.e. $v_{DD}=0$ in Eq.~\eqref{eq:q3}~\cite{Hsu2014}. This means that the effective oscillator strength
of the dipole transition vanishes and the dipolar scattering is quenched.
Second result is the formation of the high-quality resonant state due to 
the destructive interference of first and second dipole modes in the coupling to the octupole mode, i.e. $v_{DO}=0$.
It is manifested as a suppression of the radiative losses in the octupole channel,  Eq.~\eqref{eq:wO}. The effect can be also understood as a decoupling of the octupole mode from the dipole ones. The quality factor of the octupole resonance is intrinsically higher than that for the  dipole in $(\lambda/L)^{4}$ times, where $L$ is the characteristic disk size and $\lambda$ is the wavelength. However, in the general case this quality factor is suppressed due to the hybridization with the dipole mode. If there were only one dipole mode, it would hybridize with the octupole one and increase its linewidth (second term in Eq.~\eqref{eq:wO}). However, the presence of  two dipole modes enables the mutual cancellation of the radiative damping and restores the intrinsically high finesse of the octupole resonance.
 
\begin{figure}
\includegraphics[width=0.5\textwidth]{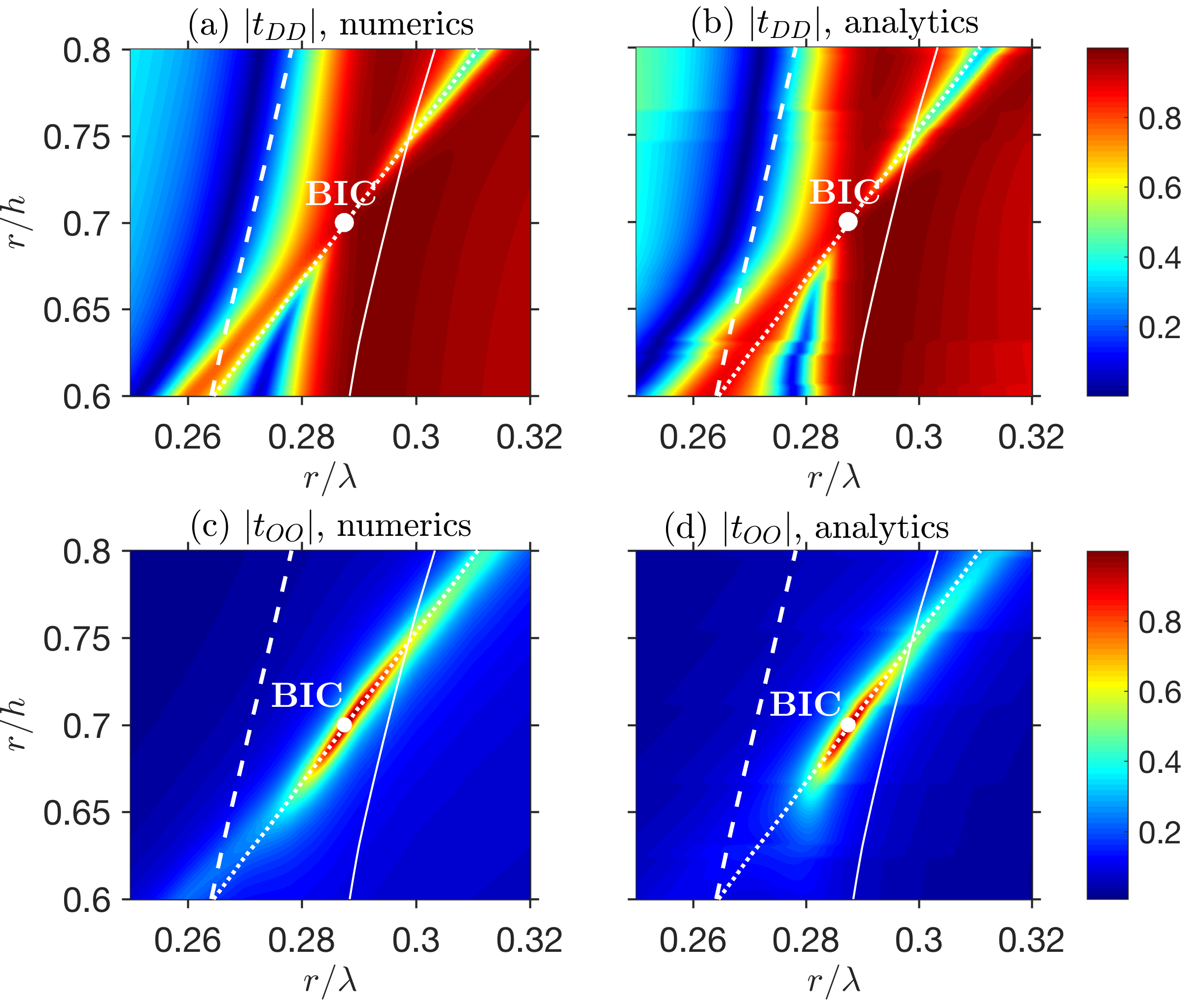}
 \caption{Comparison of the scattering coefficients in numerical (a,c) and analytical (b,d) calculations. Top and bottom panels correspond to magnetic dipole  and magnetic octupole scattering channels. Frequencies of the 
bare second  magnetic dipole mode, dark dipole state and octupole mode, used in the analytical calculation, are plotted by solid, dotted  and dashed lines, respectively. White dot  indicates the  high-$Q$ (quasi-BIC) state. The  disk permittivity is $\eps=10.73$. }\label{fig:3l}
 \end{figure}

 Figure~\ref{fig:3l} shows the dependence of the scattering spectra on the aspect ratio of the disk $r/h$. Results in the left and right columns are calculated numerically and agree with Ref.~\cite{Rybin2017}. Right column has been calculated analytically using the parameters extracted from the numerically found $Q$-matrices. We note, that in the extraction procedure it is necessary to include a small admixture of the magnetic modes with $J=5,7$ to the octupole state. This means adding a correction 
$ -Q_{\nu\nu_{1}}[Q^{-1}]_{\nu_{1}\nu_{2}}Q_{\nu_{2}\nu}$ to the reduced $2\times 2$ $Q$-matrix, where the indices $\nu,\nu'$ are in the dipole and octupole channels and the indices $\nu_{1,2}$ run over the modes with $J=5,7,\ldots$
   The calculation reveals both the dark state, where the dipole scattering is suppressed and the high-$Q$ state stemming from the octupole modes. This high-$Q$ (quasi-BIC) state is manifested as a sharp maximum  in the octupole scattering channel, 
 Fig.~\ref{fig:3l}(c,d) for $r/h\approx 0.7$ (white dot). The  dark state reveals itself as a dip in the dipole scattering channel, Fig.~\ref{fig:3l}(a,b).
An anticrossing between the dark state and the octupole resonance can be clearly seen by comparing the dip position
with the frequency of the bare dark state, corresponding to decoupled dipole and octupole (dashed line).
 To summarize, our   semi-analytical and numerical results for the linear scattering 
   are in excellent agreement, confirming our interpretation of the high-$Q$ state.
 
\section{Photon pair generation}\label{sec:generation}
  \begin{figure}[t]
  \includegraphics[width=0.5\textwidth]{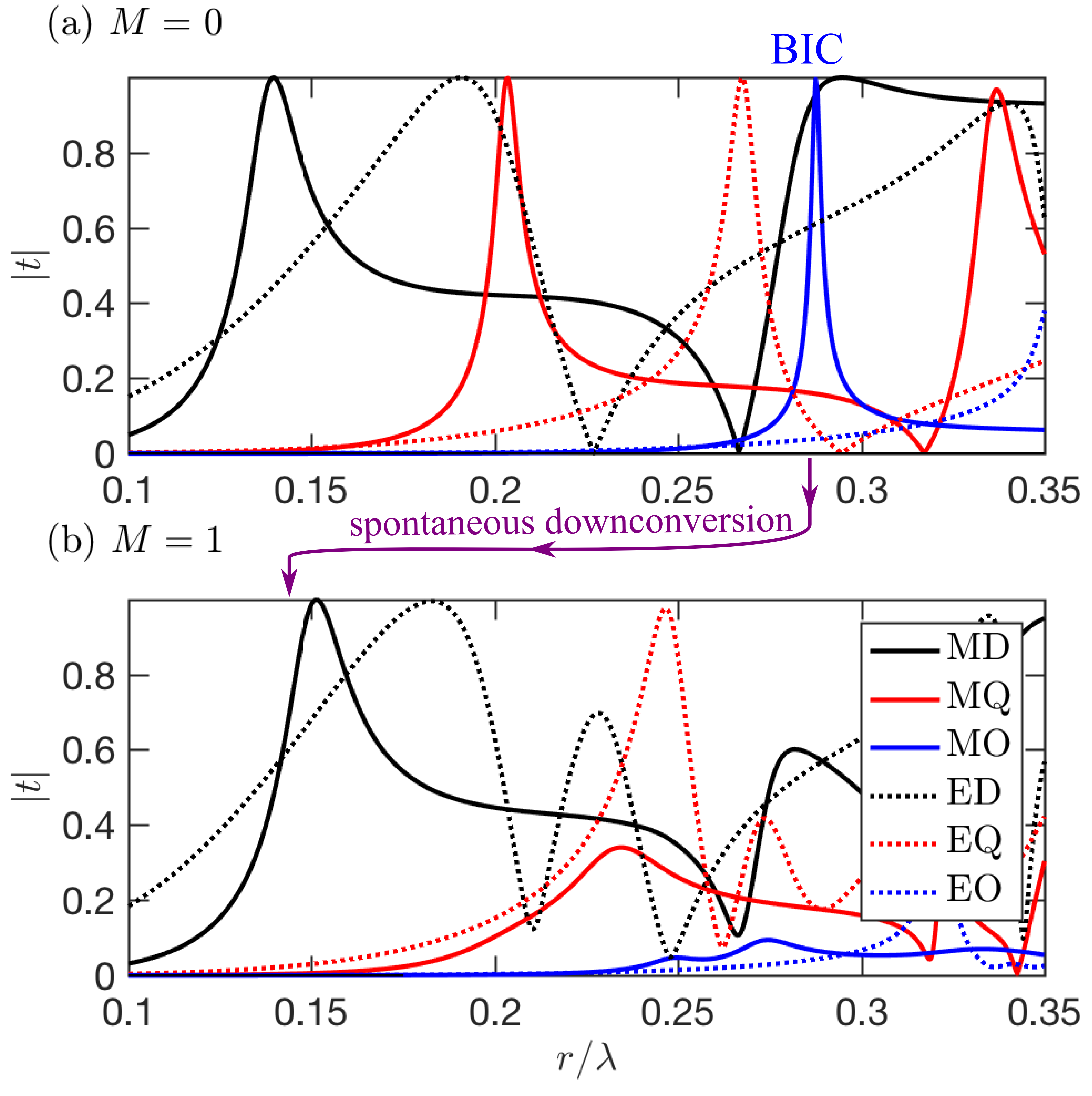}
 \caption{(a,b) Multipolar decomposition of the linear scattering spectra in the channels with $M=0$ (a) and $M=1$ (b) for $r/h=0.7$.
 The arrow indicates the process of generation of entangled electric (ED) and magnetic dipole (MD) modes due to the spontaneous decay of the high-$Q$ quasi-BIC state at $r/\lambda\approx 0.29$ indicated in (a).
 }
 \label{fig:4}
 \end{figure}
Having established a theory of the high-$Q$ state we  will now utilize the spontaneous downconversion from this state to enhance the two-photon generation.  The general expression for the quantum amplitude of the two-photon generation  
in case of stationary excitation by the monochromatic pump $\bm E_{p}(\bm r)$ at the frequency $\omega_{p}=\omega_{i}+\omega_{s}$
reads~\cite{Poddubny2016SPDC}
\begin{multline}
\psi(\bm r_{i}\alpha_{i}\omega_{i};\bm r_{s}\alpha_{s}\omega_{s})=\int\rmd^{3} r_{0} G_{\alpha_{i}\alpha_{i}'}(\omega_{i},\bm r_{i},\bm r_{0})\\\times
G_{\alpha_{s}\alpha_{s}'}(\omega_{s},\bm r_{s},\bm r_{0})\chi^{(2)}_{\alpha_{i}'\alpha_{s}'\gamma}E_{p}(\omega_{p},\bm r_{0})\:,\label{eq:psi}
\end{multline}
where $\bm r_{i,s}$, $\alpha_{i,s}$, $\omega_{i,s}$ are the coordinates, polarizations and frequencies of idler and signal photons, respectively, 
 $\chi^{(2)}$ is the nonlinear susceptibility tensor,  and $G$ is the electromagnetic Green function defined from $\rot\rot G=(\omega/c)^{2}[\eps(\bm r)G+4\pi\delta(\bm r-\bm r')]$.
  The Green function can be expressed in the basis of spherical harmonics as 
\begin{multline}
G_{\alpha\beta}(\bm r,\bm r')=4\pi\rmi k_{\rm out}^{3}\sum\limits_{\nu\nu'}
[\overline Q^{-1}]_{\nu';\nu}(-1)^{M}\\\times [\bm H_{\nu}]_{\alpha}(k_{\rm out}\bm r)
 [\overline {\bm J}_{\nu'}]_{\beta}(k_{\rm in}\bm r)\:,\label{eq:G}
\end{multline}
where the horizontal bar indicates the substitution $M\to -M$,$M'\to -M'$.
Substituting Eq.~\eqref{eq:G} into Eq.~\eqref{eq:psi} we obtain the following compact expression for the two-photon amplitude:
\begin{multline}\label{eq:final}
\psi(p\to\nu_{s}+\nu_{i})=[\overline Q^{-1}]_{\nu_{s}'\nu_{s}}(\omega_{s})[\overline Q^{-1}]_{\nu_{i}'\nu_{i}}(\omega_{i})\\\times\langle \nu_{i}',\nu_{s}'|\chi^{(2)}|\nu_{p}'\rangle [Q^{-1}]_{\nu_{p}'\nu_{p}}(\omega_{p})a_{\nu_{p}}^{(0)}\:,
\end{multline}
where $a_{\nu_{p'}}^{(0)}$ are the multipolar expansion coefficients for the incident pump wave.
The two-photon wavefunction is determined by the convolution of the inverted $Q$ matrices at pump, signal and idler frequencies weighted by the matrix elements
\begin{multline}
\langle \nu_{i}'\nu_{s}' |\chi^{(2)}|\nu_{p}'\rangle\equiv 
-(4\pi)^{2}k_{i,\rm out}^{3}k_{s,\rm out}^{3}(-1)^{M_{i}+M_{s}}\\\times
\chi_{\alpha\beta\gamma}^{(2)}\!\int\limits_{\rm (disk)}\!\rmd^{3}r[\bm {\overline J}_{\nu_{i}'}]_{\alpha}(k_{i,\rm in}\bm r)
[\bm {\overline J}_{\nu_{s}'}]_{\beta}(k_{s, \rm in}\bm r)
[\bm J_{\nu_{p}'}]_{\gamma}(k_{p,\rm in}\bm r)\:\label{eq:overlap}
\end{multline}
that describe their nonlinear coupling. Thus, the optical resonances of pump, signal and idler will be  directly manifested in the two-photon count rate. 
Equation \eqref{eq:final} rigorously incorporates the complex multipolar structure of the eigenmodes of the nanoparticle. It  generalizes the results obtained in Ref.~\cite{Olekhno2018} in the point electric dipole approximation, applicable only to deeply subwavelength sizes. 

 The general symmetry properties of the integrals \eqref{eq:overlap} will be discussed in our work Ref.~\cite{Frizyuk}  in more detail.  Next, we consider the [001]-grown AlGaAs nanodisk, with the only independent component $\chi^{(2)}_{xyz}\ne 0$~\cite{boyd2003nonlinear}.  The   spontaneous decay process, relevant for the current study, and  allowed by rotation and parity symmetry is
\begin{equation}
|{\rm MD}_{z}\rangle\to |{\rm ED}_{x},{\rm MD}_{y}\rangle+|{\rm ED}_{y},{\rm MD}_{x}\rangle\:,\label{eq:rules}
\end{equation}
where MD and ED denote magnetic and electric dipole states with the corresponding polarization. 
  \begin{figure}[t!]
  \centering \includegraphics[width=0.45\textwidth]{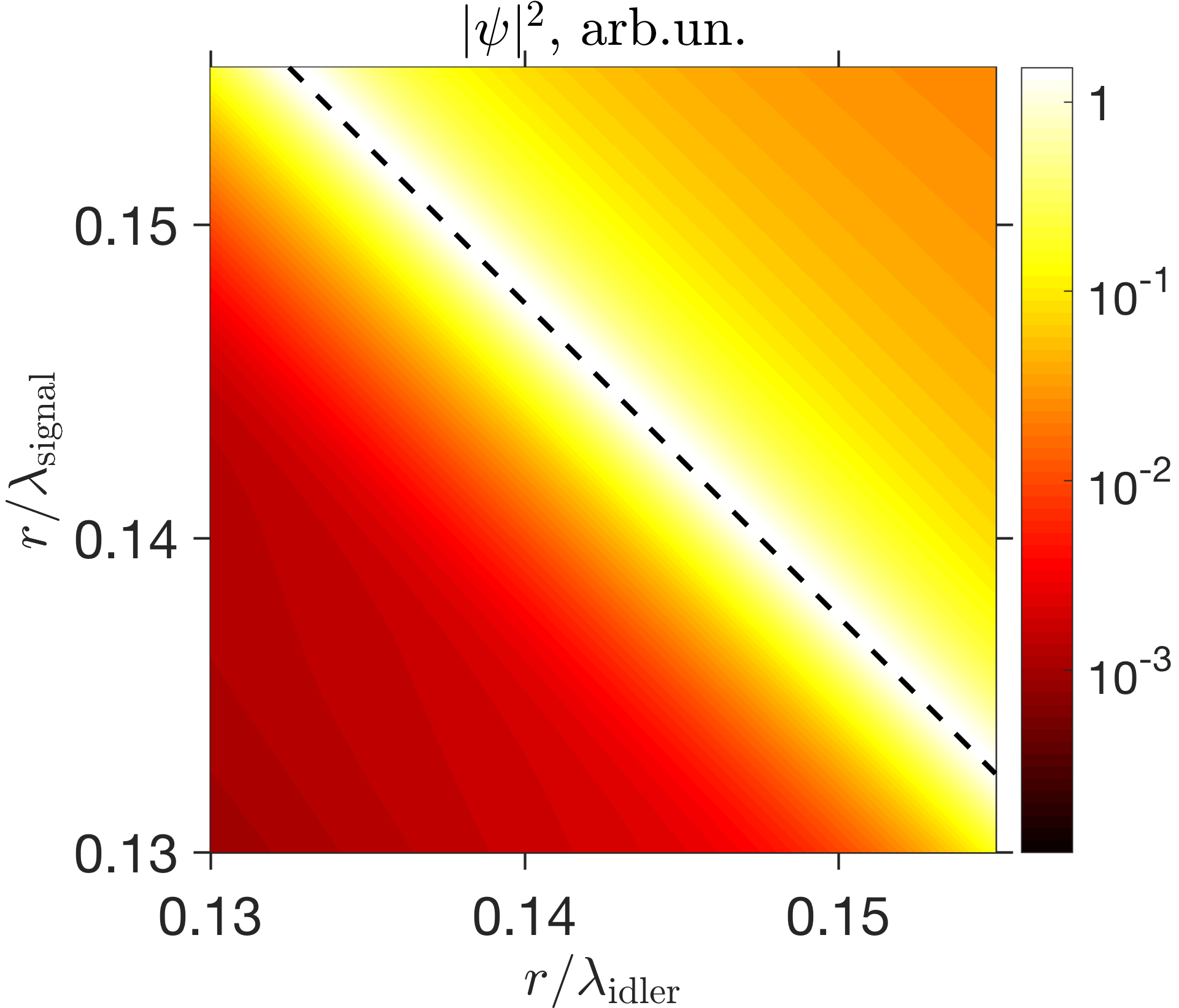}
   \caption{ (c) Square of the two-photon wavefunction depending on idler and signal photon energies.
The detection is in the magnetic dipole, $M=1$ and  electric dipole, $M=1$ signal and idler channels, respectively. Dashed line indicates the condition $r/\lambda_{i}+r/\lambda_{s}=0.29$ of the resonance with the high-$Q$ state.
Calculated for $r/h=0.7$.
} \label{fig:5}
 \end{figure}
  \begin{figure}[b]
\centering \includegraphics[width=0.45\textwidth]{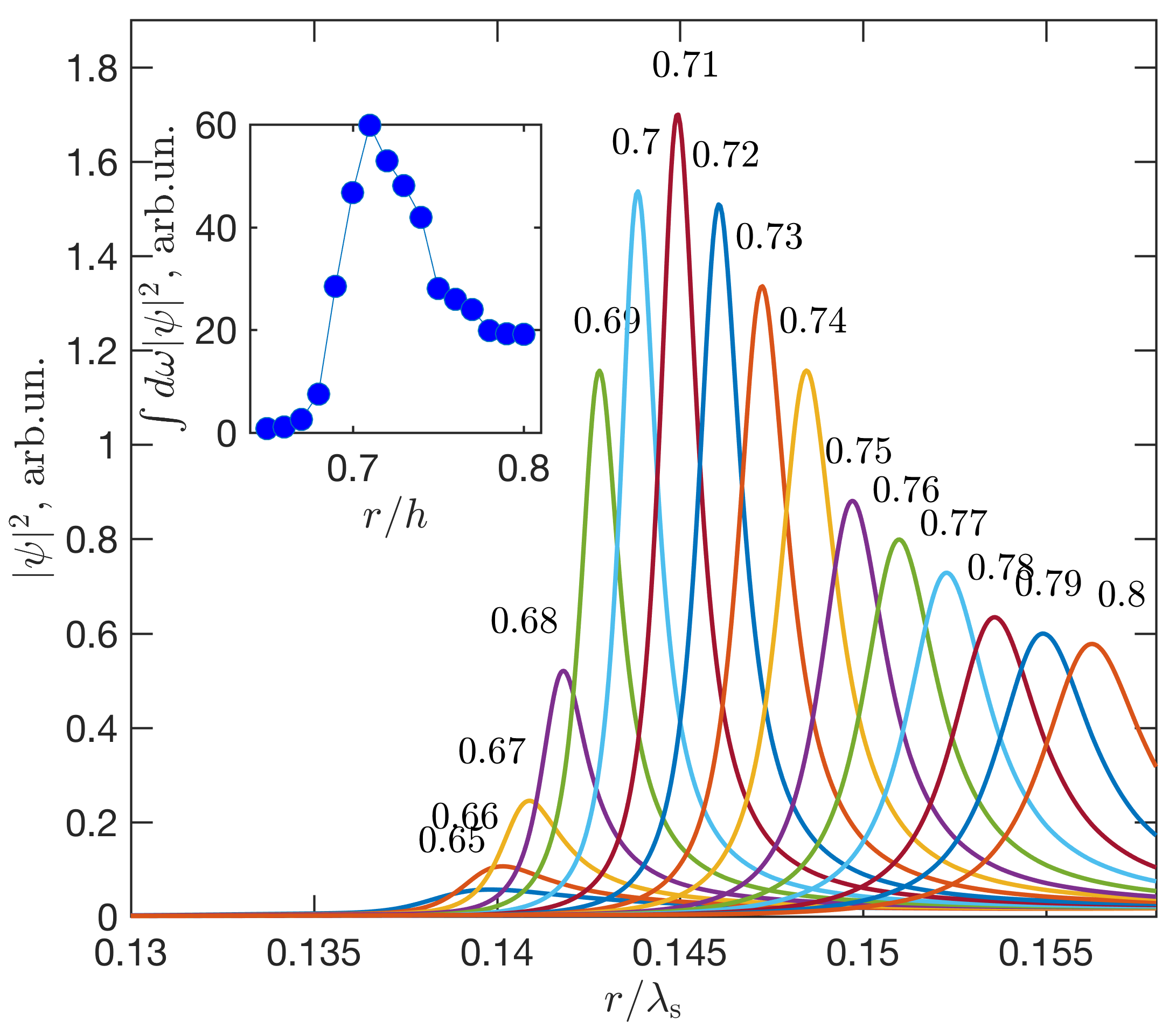}
   \caption{Spectra of degenerate SPDC ($\lambda_{i}=\lambda_{s}=2\lambda_{p}$) for different nanodisk aspect ratios $r/h$ indicated near the corresponding curves. Inset shows the dependence of  the area of the spectra on $r/h$. Other parameters are the same as in Fig.~\ref{fig:5}.} \label{fig:6}
 \end{figure}
As discussed in the previous section, the MD$_{z}$ channel has a high-Q-resonance due to the hybridization with the octupole mode. Hence, as soon as the incident pump wave excite the vertical magnetic dipole, it will couple to the high-Q-resonance and efficiently generate
entangled signal and idler photons in the ED$_{x}$ and MD$_{y}$ channels. The linear scattering 
spectra for $r/h\approx 0.7$ and the angular momentum projections $M=0,1$ are shown in Fig.~\ref{fig:4}(a,b), respectively. They confirm that the spontaneous parametric decay of the high-$Q$ state at the frequency $r/\lambda\approx 0.3$ [Fig.~\ref{fig:4}(a)] will lead to the generation of the in-plane signal and idler electric and magnetic dipole modes with $r/\lambda\approx 0.15$ [Fig.~\ref{fig:4}(b)]. The relevant process is indicated by a violet arrow.
In our calculation we excite the structure obliquely under the angle $\theta=45^{\circ}$ from the $z$-axis by a TE-polarized plane wave. The resulting square of the two-photon wavefunction is shown in Fig.~\ref{fig:5}.
It demonstrates a sharp maximum when the sum of signal and idler frequencies matches the high-$Q$ state frequency,
$r/\lambda_{i}+r/\lambda_{s}= 0.2875$ (see the dashed line). 
In order to further confirm that the maximum in SPDC is related to the formation of  high-$Q$ state we examine the dependence of the SPDC spectra on the nanodisk aspect ratio. Figure~\ref{fig:6} presents the dependence of
the degenerate  SPDC with $\lambda_{i}=\lambda_{s}$ on $r/h$. The spectra feature a prominent maximum that becomes sharper and higher when the aspect ratio is tuned to the value $r/h=0.71$. This is in full agreement with the formation of high-Q state in the  linear scattering  spectra, see Fig.~\ref{fig:3l}. The asymmetry in the behavior on the peak maximum with $r/h$ traces the asymmetric profile of the MD mode at the pump frequency (black curve in Fig.~\ref{fig:4}a).
 The spectrally integrated degenerate SPDC efficiency increases near $r/h=0.71$ as well, see the inset of Fig.~\ref{fig:6}.

\section{Conclusion}
To conclude, we have proposed the high-quality states (quasi-bound states in continuum) in AlGaAs nanodisks for the spontaneous nonlinear generation of quantum-entangled photon pairs. 
Specifically, we have considered a process when the TE-polarized plane wave pump is obliquely incident at the frequency in the vicinity of the high-$Q$ mode and the signal and idler photons are generated in the in-plane electric and magnetic dipole modes. For a certain aspect ratio of a disk, corresponding to the formation of high-$Q$-state, the generation of entangled pairs is resonantly enhanced. Our results are supported by rigorous full-wave numerical  modeling, symmetry analysis and analytical theory, revealing the multipolar nature of the high-$Q$ state.


%

\end{document}